\documentclass[letterpaper]{article} % DO NOT CHANGE THIS
\usepackage{aaai25}  % DO NOT CHANGE THIS
\usepackage{times}  % DO NOT CHANGE THIS
\usepackage{helvet}  % DO NOT CHANGE THIS
\usepackage{courier}  % DO NOT CHANGE THIS
\usepackage[hyphens]{url}  % DO NOT CHANGE THIS
\usepackage{graphicx} % DO NOT CHANGE THIS
\usepackage{natbib}  % DO NOT CHANGE THIS AND DO NOT ADD ANY OPTIONS TO IT
\usepackage{caption} % DO NOT CHANGE THIS AND DO NOT ADD ANY OPTIONS TO IT

\usepackage{amsmath}
\usepackage{amssymb}
\usepackage{graphicx}
\usepackage{pdfpages}
\usepackage{multirow}
\usepackage{booktabs}

\usepackage{algorithm}
\usepackage{algorithmic}

\usepackage{newfloat}
\usepackage{listings}
\DeclareCaptionStyle{ruled}{labelfont=normalfont,labelsep=colon,strut=off} % DO NOT CHANGE THIS
\floatstyle{ruled}
\newfloat{listing}{tb}{lst}{}
\floatname{listing}{Listing}
\newcommand{\figref}[1]{Figure~\ref{fig:#1}}
\newcommand{\tabref}[1]{Table~\ref{tab:#1}}

\title{RDSinger: Reference-based Diffusion Network for Singing Voice Synthesis}
\author{
    Kehan Sui\textsuperscript{\rm 1}, Jinxu Xiang\textsuperscript{\rm 2},
    Fang Jin\textsuperscript{\rm 1}
}

\affiliations{
    \textsuperscript{\rm 1}George Washington University\\
    \textsuperscript{\rm 2}Platform and Content Griuo (PCG), Tencent \\
    suikehan@gwu.edu, jinxuxiang@global.tencent.com, fangjin@gwu.edu
}

\usepackage{bibentry}
\begin{document}

\maketitle

\begin{abstract}
Singing voice synthesis (SVS) aims to produce high-fidelity singing audio from music scores, requiring a detailed understanding of notes, pitch, and duration, unlike text-to-speech tasks. Although diffusion models have shown exceptional performance in various generative tasks like image and video creation, their application in SVS is hindered by time complexity and the challenge of capturing acoustic features, particularly during pitch transitions. Some networks learn from the prior distribution and use the compressed latent state as a better start in the diffusion model, but the denoising step doesn't consistently improve quality over the entire duration. We introduce RDSinger, a reference-based denoising diffusion network that generates high-quality audio for SVS tasks. Our approach is inspired by Animate Anyone, a diffusion image network that maintains intricate appearance features from reference images. RDSinger utilizes FastSpeech2 mel-spectrogram as a reference to mitigate denoising step artifacts. Additionally, existing models could be influenced by misleading information on the compressed latent state during pitch transitions. We address this issue by applying Gaussian blur on partial reference mel-spectrogram and adjusting loss weights in these regions. Extensive ablation studies demonstrate the efficiency of our method. Evaluations on OpenCpop, a Chinese singing dataset, show that RDSinger outperforms current state-of-the-art SVS methods in performance.
\end{abstract}

\begin{figure*}[t]
    \centering
    \includegraphics[width=\hsize]{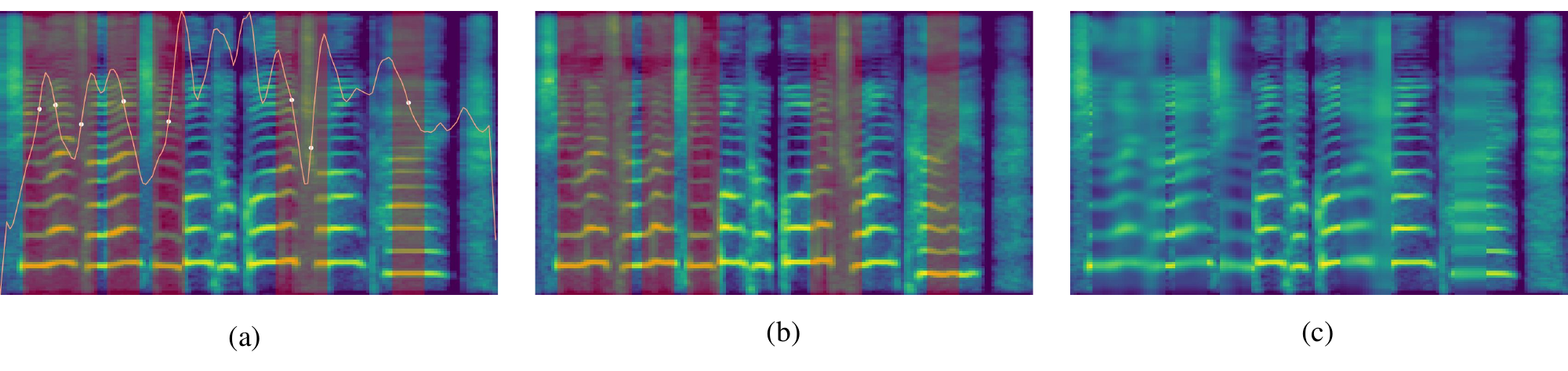}
    \caption{The figure illustrates the pitch transition region. Panel (a) shows the pitch transition generated by FastSpeech2, with the orange line representing the high-to-low frequency ratio and white dots marking transition points. The red area highlights the transition zone. Panel (b) shows the alignment of FastSpeech2's pitch transition region with the ground-truth mel-spectrogram, indicating that the transition regions correspond to areas that need enhancement. Panel (c) displays the result of applying Gaussian blur to the pitch transition region shown in (a).
}
    \label{fig:pitch}
\end{figure*}

\section{Introduction}
Audio processing encompasses several active research areas: \cite{popov2021grad} \cite{oord2016wavenet} \cite{kim2021conditional} are the popular methods in speech synthesis, \cite{roberts2018hierarchical} \cite{dong2018musegan} have accomplished significant result in music synthesis, and \cite{defossez2020real} \cite{xiang2022dynamic} have achieved the real-time audio denoising. Significant advancements have been made in automatic voice generation through generative network models. Text-to-speech (TTS) systems now produce high-quality, human-like audio, applicable in various domains such as video editing and audiobook production. While singing voice synthesis (SVS) shares similar objectives with TTS, SVS focuses on generating high-quality singing audio from lyrics and musical scores. In addition to the requirements of speech synthesis, SVS must capture acoustic features like pitch and duration, which are critical for singing. The typical SVS pipeline involves two main stages: an acoustic model that generates intermediate outputs with acoustic features such as mel-spectrograms \cite{shen2018natural}, and a vocoder that transforms these intermediate outputs into waveforms. 

Commonly used vocoders, such as Parallel WaveGAN \cite{yamamoto2020parallel} and HiFi-GAN \cite{kong2020hifi}, deliver high-fidelity performance. Existing work on acoustic models could be roughly divided into two categories. One is the generative adversarial network (GAN) based model \cite{goodfellow2020generative}. These models typically use a generator to reconstruct acoustic features from lyrics and music scores, while the discriminator differentiates between real and generated acoustic features. However, these networks have limited reproducibility due to the stochastic training process and sensitivity to initialization. Another type of acoustic model is the diffusion-based network. The diffusion probabilistic model \cite{ho2020denoising} has shown remarkable performance in image generation \cite{rombach2022high} and video generation \cite{guo2023animatediff} tasks. Unlike GAN networks, diffusion models could be stably trained but are challenged by their computational complexity. HiddenSinger \cite{hwang2023hiddensinger} addresses this problem by sampling a latent representation from a musical score. DiffSinger \cite{liu2022diffsinger} proposed a shallow diffusion network to speed up inference. It substitutes the initial Gaussian noise with a diffused mel-spectrogram to make better use of the prior knowledge. 

Diffusion-based networks have successfully addressed issues related to training stability and have demonstrated efficiency in the inference process. These networks typically rely on an encoded embedding derived from the music score to condition the denoising process. However, this embedding may not encapsulate sufficient information to effectively guide the denoising across all timeframes. We observed that in certain segments, the denoised mel-spectrogram does not exhibit improvements in quality compared to the intermediate mel-spectrogram generated during earlier stages of the process. This limitation becomes particularly evident when examining the synthesized audio produced by other models, where audible artifacts such as glitches and cracks appear, especially during the pitch transition regions. These regions include critical sections of the singing voice, such as note terminations, phoneme transitions, and pitch shifts, where the models struggle to maintain smooth and natural output. These imperfections highlight the challenges associated with ensuring that the conditioning mechanism in diffusion-based networks consistently guides the denoising process to produce high-quality, artifact-free audio across the entire duration of the synthesized output.

In this work, we introduce RDSinger, a diffusion-based acoustic model that generates high-fidelity singing by transforming noise into mel-spectrograms, conditioned on the music score and a reference mel-spectrogram. We adapt the network design from DiffSinger and Animate Anyone \cite{hu2024animate}, modifying the audio-denoising network to incorporate multiple conditions, enhancing guidance during the denoising process. While our model accepts both the music score embedding and the reference mel-spectrogram as input, we can comprehensively learn the relationship with the reference mel-spectrogram in a consistent feature space, which significantly contributes to the improvement of audio quality. 

Most SVS networks use the L1 or L2 loss function across the entire mel-spectrogram, assuming each pixel contributes equally to the overall loss. However, this assumption may overlook the fact that different regions of the spectrogram could have varying levels of significance in influencing the perceptual quality of the synthesized audio. To address this, we applied a Gaussian blur to the regions of the mel-spectrogram that might introduce artifacts or degrade audio quality if not handled properly. This adjustment was intended to reflect the varying importance of different regions in the spectrogram by emphasizing those areas that are more critical to the synthesis process. Consequently, we modified the loss function to apply differential weighting, thereby directing greater focus and computational resources to these strategically identified regions. Through the combination of these refined techniques—targeted Gaussian blurring and adjusted loss weighting—our proposed model, RDSinger, is capable of generating audio that is not only of high technical quality but also exhibits a significantly more natural and lifelike vocal timbre. These enhancements collectively contribute to the superior performance of RDSinger in producing realistic singing voices, highlighting the effectiveness of our approach in overcoming the limitations of traditional SVS loss functions. 

\begin{figure*}[t]
    \centering
    \includegraphics[width=\hsize]{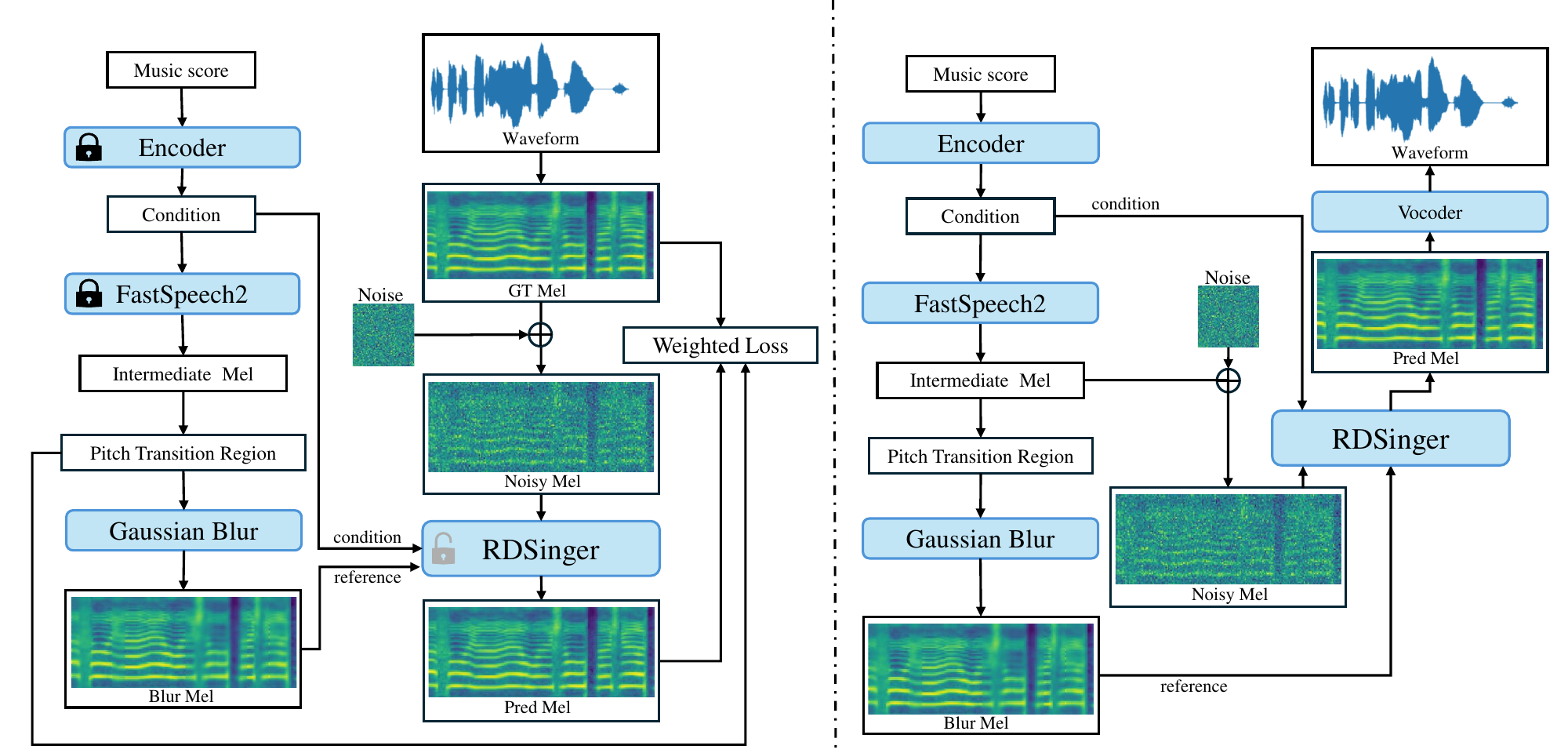}
    \caption{The training (left) and inference (right) process of RDSinger}
    \label{fig:model}
\end{figure*}

\section{Related Work}
\subsection{Neural Speech Processing}
Neural Speech Processing is an advanced field of artificial intelligence that focuses on the use of neural networks to analyze, understand, and generate human speech. It encompasses a wide range of applications, including Automatic Speech Recognition (ASR), audio denoising and Text-to-Speech (TTS), among others. Traditional speech processing methods relied heavily on handcrafted features and statistical methods, but neural speech processing leverages deep learning to learn features directly from raw data, leading to improvements in accuracy and performance. In ASR, neural networks convert spoken language into text by recognizing patterns in the audio signals. These systems have evolved from simple feedforward networks to more sophisticated architectures like transformers, which excel at handling sequential data. Audio denoising with deep learning involves using neural networks like autoencoders to reduce noise in audio signals. These models learn to differentiate between noise and the desired signal by training on large datasets. Deep learning based denoising outperforms traditional methods, providing clearer, more natural audio, and is widely used in applications like voice enhancement, communication systems, and music production. TTS synthesis, another key area within neural speech processing, involves converting written text into spoken language using speech synthesis technologies underpinned by deep learning models. Notable TTS systems such as FastSpeech2 \cite{ren2020fastspeech}, Tacotron2 \cite{shen2018natural}, and WaveNet are capable of accurately capturing prosody, pitch, duration, and energy patterns within speech. These models generate two-dimensional mel-spectrograms as intermediate outputs, which are then converted into waveform audio using vocoder networks. Vocoders like HiFi-GAN and WaveGlow \cite{prenger2019waveglow} have demonstrated exceptional performance in translating mel-spectrograms into high-quality, natural-sounding audio. 

\subsection{Singing Voice synthesis}
Text-to-Speech (TTS) converts written text into spoken audio, focusing on generating natural-sounding speech from any text input. Singing Voice Synthesis (SVS), on the other hand, generates singing voice from text and musical notes. SVS models not only convert text to audio but also capture the melody, pitch, and rhythm associated with singing. SVS is more complex due to the need to accurately reproduce musical nuances and is used in applications like virtual singers and music production. Similar with TTS, traditional SVS networks also follow the 2-step procedure. To reduce the information gap between the two steps, VISinger \cite{zhang2022visinger} and WaveGAN \cite{pena2021wave} provide the end-to-end SVS network that can model the rich acoustic variation in singing, without the need for an intermediate representation like a spectrogram. However, these GAN-based network have limitations in training stability. As the diffusion probabilistic model becomes more popular and shows an outcoming result in generation tasks, DiffSinger and HiddenSinger proposed the diffusion-based network SVS network, and DiffSinger also proposed a shallow diffusion to make better use of the prior knowledge so that it requires fewer steps in the inference procedure.

\subsection{Diffusion Probabilistic Model}
Diffusion models are versatile generative models that have found applications across various fields due to their ability to produce high-quality and realistic data. In image generation, they are used to create photorealistic images with intricate details \cite{dhariwal2021diffusion} \cite{ramesh2022hierarchical}, often outperforming traditional methods like GANs. In video generation, diffusion models help maintain temporal coherence, making them ideal for tasks like video synthesis, animation, and video super-resolution \cite{guo2023animatediff} \cite{blattmann2023align}. Beyond visual applications, these models are used in TTS synthesis, producing natural-sounding speech with fewer distortions. Grad-TTS \cite{popov2021grad} and DiffWave \cite{kong2020diffwave} have demonstrated that diffusion models perform effectively in capturing acoustic features from music scores, showcasing strong capabilities in this area. In audio processing, they improve denoising and enhance the quality of audio signals. Their flexibility and high fidelity make diffusion models valuable in diverse applications, from creative industries to scientific research.

\begin{figure}[t]
    \centering
    \includegraphics[width=\hsize]{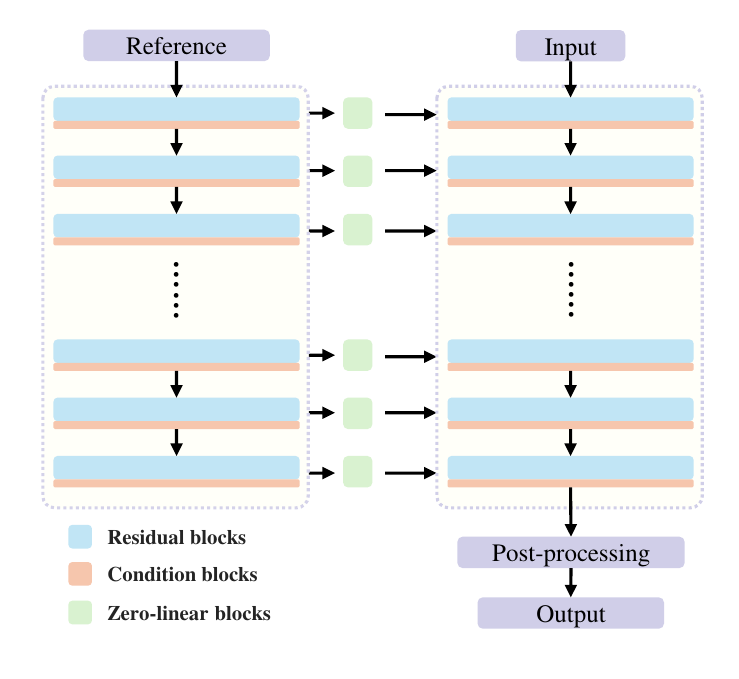}
    \caption{The diffusion network structure of RDSinger}
    \label{fig:net}
\end{figure}

\section{Methodology}

\subsection{Diffusion Model}
A diffusion model operates through two primary phases: the forward and reverse processes. In the forward process, noise is progressively added to data through multiple iterations, corrupting it until it resembles pure random noise. This progression is typically modeled as a Markov chain, where each step adds a small amount of noise, resulting in a smooth transition from clean input to noisy output. The reverse process aims to restore the original data by systematically removing noise from the corrupted input. Beginning with a noise-infused input, the model applies learned transformations step by step to gradually eliminate the noise. This iterative approach efficiently reconstructs high-quality outputs, by reversing the noise accumulation from the forward process. In recent developments, deep neural networks are widely used to handle these reverse tasks.

\subsubsection{Diffusion Process}
Define the data distribution as $p(x)$ where $x_0 \sim p(x)$ is a clean sample randomly drawn from the distribution. The diffusion process or forward process is a fixed Markov chain that gradually adds Gaussian noise to the data according to a variance schedule $\beta_1, ..., \beta_T$:
\[
q(\mathbf{x}_t | \mathbf{x}_{t-1}, \mathbf{c}) = \mathcal{N}(\mathbf{x}_t; \sqrt{1-\beta_t} \mathbf{x}_{t-1}, \beta_t \mathbf{I})
\]
where \( \mathbf{c} \) represents the condition and the posterior distribution of $q(x_{1:T}|x_0)$ could be written as:
$$q(x_{1:T}|x_0,\mathbf{c}) = \prod_{t=1}^{T}q(x_t|x_{t-1},\mathbf{c})$$
We can derive the distribution of \(\mathbf{x}_t\) directly from \(\mathbf{x}_0\) by recursively applying the Gaussian distribution from step 1 to step \(T\):
\begin{equation}
q(\mathbf{x}_t | \mathbf{x}_0,\mathbf{c}) = \mathcal{N}(\mathbf{x}_t; \sqrt{\bar{\alpha}_t} \mathbf{x}_0, (1 - \bar{\alpha}_t) \mathbf{I})
\end{equation}
where \(\bar{\alpha}_t = \prod_{i=1}^{t} (1 - \beta_i)\). This equation shows that \(\mathbf{x}_t\) can be sampled directly from \(\mathbf{x}_0\) by scaling \(\mathbf{x}_0\) and adding Gaussian noise.
\subsubsection{Reverse Process}
The reverse process aims to reverse the forward diffusion by gradually removing the noise added at each step, it is defined as a series of conditional Gaussian distributions:
\[
p_\theta(\mathbf{x}_{t-1} | \mathbf{x}_t, \mathbf{c}) = \mathcal{N}(\mathbf{x}_{t-1}; \mu_\theta(\mathbf{x}_t, t, \mathbf{c}), \Sigma_\theta(\mathbf{x}_t, t, \mathbf{c}))
\]
where \( \mu_\theta(\mathbf{x}_t, t, \mathbf{c}) \) and \( \Sigma_\theta(\mathbf{x}_t, t, \mathbf{c}) \) are the mean and variance parameters learned by the model.
\newline
The mean \( \mu_\theta(\mathbf{x}_t, t, \mathbf{c}) \) in the reverse process can be derived as:
\[
\mu_\theta(\mathbf{x}_t, t, \mathbf{c}) = \frac{1}{\sqrt{\alpha_t}} \left( \mathbf{x}_t - \frac{\beta_t}{\sqrt{1 - \bar{\alpha}_t}} \epsilon_\theta(\mathbf{x}_t, t, \mathbf{c}) \right)
\]
where \( \beta_t \) is the noise schedule parameter and \( \epsilon_\theta(\mathbf{x}_t, t, \mathbf{c}) \) is the neural network's prediction of the noise component at time \( t \).
\newline
The variance \( \Sigma_\theta(\mathbf{x}_t, t, \mathbf{c}) \) can be modeled as:
\[
\Sigma_\theta(\mathbf{x}_t, t, \mathbf{c}) = \beta_t \mathbf{I}
\]

\subsubsection{Training the Model}
The model is trained by minimizing the expected mean squared error between the true noise \( \epsilon \) and the predicted noise \( \epsilon_\theta(\mathbf{x}_t, t, \mathbf{c}) \). The loss function for training is given by:
\[
L_{\text{simple}} = \mathbb{E}_{\mathbf{x}_0, t, \epsilon, \mathbf{c}} \left[ \|\epsilon - \epsilon_\theta(\mathbf{x}_t, t, \mathbf{c})\|^2 \right]
\]
where \( \epsilon \) is the true noise that was added in the forward process, \( \mathbf{x}_0 \) is the original data, and \( \mathbf{x}_t \) is the noisy data at time step \( t \), which can be computed as:
    \[
    \mathbf{x}_t = \sqrt{\bar{\alpha}_t} \mathbf{x}_0 + \sqrt{1 - \bar{\alpha}_t} \epsilon
    \]

\begin{figure*}[t]
    \centering
    \includegraphics[width=\hsize]{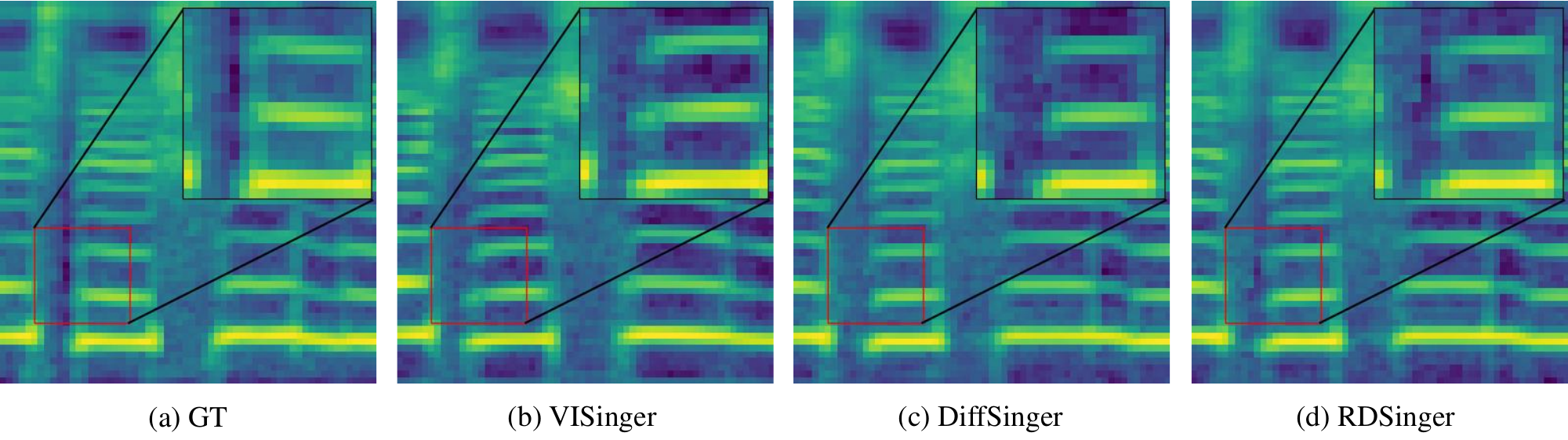}
    \caption{Visualization of generated samples with varying systems: (a) GT, (b) VISinger, (c) DiffSinger, (d)RDSinger.}
    \label{fig:4mel}
\end{figure*}

\subsection{Reference-based Diffusion}
Diffusion models generate results that are correlated with the conditioning content. DiffSinger employs a shallow diffusion model, which introduces noise to the intermediate mel-spectrogram generated by other models to guide the diffusion process. However, this method struggles to preserve fine details, a challenge addressed by utilizing a reference network. The concept of integrating a reference network into diffusion models was first introduced in AnimateAnyone \cite{hu2024animate}. The architecture of the reference network mirrors that of the denoising network. The reference network benefits from the pre-trained spectrogram feature extraction of the original model, leading to a well-initialized feature set. Additionally, due to the similar structure and shared initial weights with the denoising network, the denoising network can selectively learn correlated features from the reference network within the same feature space.

In RDSinger, our diffusion network contains two main components: a reference network and a denoising network, as illustrated in \figref{net}. The mel-spectrogram generated by FastSpeed2 is fed into the reference network. During the processing, the hidden output from each layer in the reference network is added to the corresponding layer output in the denoising network. We apply the trainable zero-linear layer on the reference hidden states. The zero-linear layer is a simple fully connected layer in which both the weight and bias are initialized to zero, ensuring that the reference hidden state values added to the denoising network start as zeros during the initial phase of training. The pipeline of our method is illustrated in \figref{model}.

\subsection{Pitch Transition Enhancement}
The quality of generated audio in SVS tasks remains inadequate in current models, particularly during pitch transition regions, where noticeable artifacts can be frequently heard. RDSinger addresses this issue by adjusting the training strategy to apply varying loss weights across different regions, assigning higher weights specifically to pitch transition regions. To identify the pitch transition region from a mel-spectrogram \( \mathbf{M} \) with dimensions \(F \times T \), where \( F \) is the number of frequency bins, and \( T \) is the number of time steps, we first split the frequency dimension into two halves: low-frequency energy: \( E_{\text{low}}(t) \) and high-frequency energy: \( E_{\text{high}}(t) \).

The energy in each region is computed as:
\[
E_{\text{low}}(t) = \sum_{f=1}^{\frac{F}{2}} \mathbf{M}(f, t) \quad E_{\text{high}}(t) = \sum_{f=\frac{F}{2}+1}^{F} \mathbf{M}(f, t)
\]
The ratio of high-frequency to low-frequency energy is then given by:
\[
R(t) = \frac{E_{\text{high}}(t)}{E_{\text{low}}(t) + \epsilon}
\]
where \( \epsilon \) is a small constant (e.g., \( 10^{-6} \)) added to prevent division by zero.
To smooth the ratio \( R(t) \), apply a 1D convolution with a uniform kernel of size \( k \):

\[
R_{\text{smooth}}(t) = \frac{1}{k} \sum_{i=-\frac{k-1}{2}}^{\frac{k-1}{2}} R(t+i)
\]
\newline
To find pitch transition points, first compute the mean of the smoothed ratio $\bar{R}_{\text{smooth}}$ by taking the mean on $R_{\text{smooth}}(t)$. Then, identify points where the sign of \( R_{\text{smooth}}(t) - \bar{R}_{\text{smooth}} \) changes:
\[
\text{S}(t) = \text{sign}\left(R_{\text{smooth}}(t) - \bar{R}_{\text{smooth}} \right)
\]
Transition points are the indices where the sign of \( \text{S}(t) \) differs from \( \text{S}(t-1) \). For each transition point \( t_s \), define a window of size \( w \) around it to be the regions of pitch transitions.
 
Specifically, to mitigate distortion in the pitch transition regions of the audio, we compute adjustments based on the areas surrounding the pitch transition points over the whole mel-spectrogram timeframe. We applied a Gaussian blur with a window size of 5 inside each transition region, and the resulting mel-spectrogram is shown in \figref{pitch}. The application of Gaussian blur softens edges and diminishes detail, which helps to mitigate sharp transitions in spectrogram regions that may contain erroneous reference information. During the training of the diffusion model, we assign greater weight to these regions in the loss computation. This selective exposure and weighting process directs the model's learning to focus on the important regions, which are crucial for auditory perception. Consequently, this approach effectively reduces distortion and enhances the overall realism of the generated content.

\begin{table*}[h]
\centering
\begin{tabular}{l|cccc}
\toprule[2pt]
Method & \qquad Ranking (↓)  \qquad\quad & \qquad MOS (↑)\qquad\quad  & \qquad SIG MOS (↑)\qquad\quad  & \qquad BAK MOS (↑)\qquad\quad \\
\hline
GT & 1.18 ± 0.04 & 4.61 ± 0.04 & 4.15 ± 0.04 & 4.33 ± 0.04 \\
GT-Mel-HiFiGAN & 1.85 ± 0.04 & 4.43 ± 0.04 & 3.87 ± 0.05 & 4.27 ± 0.04 \\
\hline
NaiveRNN & 6.49 ± 0.11 & 2.91 ± 0.07 & 3.13 ± 0.05 & 3.63 ± 0.07 \\
XiaoiceSing & 6.64 ± 0.12 & 2.87 ± 0.09 & 3.24 ± 0.07 & 3.58 ± 0.08 \\
FastSpeech2 & 5.65 ± 0.15 & 3.11 ± 0.07 & 3.21 ± 0.06 & 3.79 ± 0.07 \\
VISinger & 4.66 ± 0.13 & 3.43 ± 0.08 & \textbf{3.47} ± 0.06 & 3.64 ± 0.07 \\
Diffsinger & 5.09 ± 0.11 & 3.30 ± 0.07 & 3.39 ± 0.05 & 3.98 ± 0.06 \\
RDSinger(Ours) & \textbf{4.45} ± 0.12 & \textbf{3.46} ± 0.07 & 3.43 ± 0.06 & \textbf{4.07} ± 0.06 \\
\bottomrule[2pt]
\end{tabular}
\caption{Integrated inference results of SVS tasks}
\label{tab:exp:main}
\end{table*}

\section{Experiments}
This section begins with a detailed description of the experimental setup. Subsequently, it presents and thoroughly analyzes the primary results for the SVS task. Ablation studies are then conducted to evaluate the contribution of each component to the model's overall performance.

\subsection{Experimental Setup}
\subsubsection{Dataset}
The Opencpop dataset \cite{wang2022opencpop} is a publicly available, high-quality Mandarin singing corpus designed for SVS systems. It comprises 100 distinct Mandarin songs, all performed by a professional female vocalist. The recordings were produced in a controlled, professional studio environment with a sampling rate of 44.1KHz. The dataset is annotated with detailed phonetic information, including syllables, note pitch, duration, and phonemes. It contains a total of approximately 5.2 hours of audio content. The raw data contains 100 audios, and we use the officially launched TextGrid to segment the original audios which results in 3756 utterance segments in total, each piece is within the time of 5 to 15 seconds. For validation and testing purposes, we randomly select 3 songs, which yields 126 utterance segments.

\subsubsection{Data Preprocessing}
The mel-spectrogram is computed as the target ground truth for RDSinger using Short-Time Fourier Transform (STFT) with a hop size of 128 and a frame size of 512. The dimension of mel-spectrogram is set to be 80. Meanwhile, we also use Parselmouth \cite{parselmouth} to extract the ground-truth F0 following \cite{wu2020adversarially} and the F0 is normalized to have zero mean and unit variance. 

\subsubsection{Implementation Details}
To condition the model, several encoders are employed. Initially, the PyPinyin library \cite{ren2020deepsinger} is used to decompose Chinese lyrics into individual syllables, which improves the network’s recognition accuracy. The lyrics encoder then transforms these phoneme IDs into embedding sequences and processes them into phoneme hidden sequences through multiple transformer blocks. A length regulator subsequently adjusts the phoneme-level representation to match the frame-level representation based on the ground-truth duration. Both the phoneme and note pitch embeddings are set to a dimension of 256. The note pitch encoder, following the approach in FastSpeech2, converts the note pitch into a pitch contour. Finally, the phoneme and pitch representations are combined to serve as the input condition for the network. We use FastSpeed2 to generate the intermediate mel-spectrogram and add Gaussian blur during pitch transition regions. In the diffusion network, the blurred mel-spectrogram from FastSpeech2 serves as the reference.

\subsubsection{Training Setup}
Both the denoising network and the reference network in RDSinger are initialized using the a training checkpoint from DiffSinger. During our training, the weight of each network component are updated independently. As shown in \figref{model}, we use a pre-trained music score encoder and FastSpeech2 to generate an intermediate reference mel-spectrogram from the music scores. In our work, only the diffusion network is trained. RDSinger was trained on a single NVIDIA 4090 GPU with a batch size of 48 for approximately 12 hours, utilizing the Adam optimizer with a learning rate of $lr = 10^{-4}$.

\subsection{Main Results and Analysis}

\subsubsection{Quantitative Metrics}
To evaluate the performance of different models, we consistently employed a pre-trained HiFiGAN vocoder to convert the generated mel-spectrogram into waveform. The evaluation encompassed the following methods: 1) GT, ground-truth audio, 2) GT-Mel-HiFiGAN: ground-truth audio converted to mel-spectrogram and then back to audio using the HiFiGAN vocoder, 3) NaiveRNN, the RNN-based SVS network proposed by \cite{watanabe2018espnet}, 4) XiaoiceSing, a GAN-based SVS network utilizing the FFT block as the main model structure, 5) FastSpeech2, an end-to-end TTS model incorporating a duration predictor and a pitch predictor, enabling SVS tasks, 6) VISinger, a state-of-the-art GAN-based end-to-end SVS system, 7) DiffSinger, a diffusion-based SVS network which uses shallow diffusion approach to speed up inference, 8)RD-Singer, our proposed model. For the diffusion-based models RDSinger and DiffSinger, we uniformly use 100 denoising steps to evaluate metrics as presented in \tabref{exp:main}.

\begin{figure}[t]
    \centering
    \includegraphics[width=1.0\hsize]{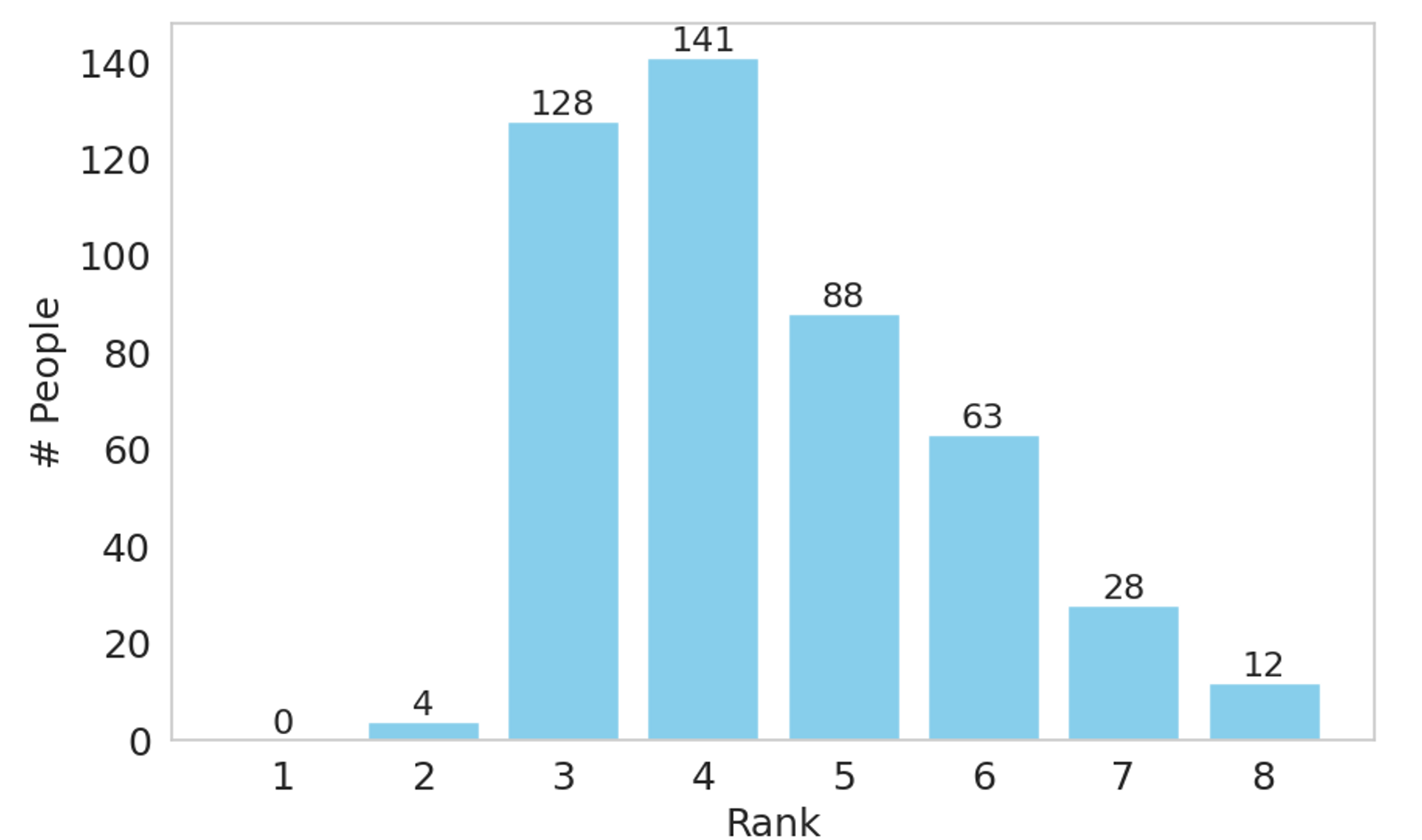}
    \caption{Distribution of Ranking Scores for RDSinger}
    \label{fig:people}
\end{figure}

Evaluating models for SVS tasks is relatively challenging because the generated audio often depends on the duration of the given phonemes or units, leading to discrepancies between the length of the generated and ground-truth audio. As a result, traditional audio metrics, such as PESQ \cite{rix2001perceptual} and STOI \cite{taal2011algorithm}, are ineffective for making valid comparisons.

To assess perceptual audio quality, we employ two types of metrics. The first type is human testing, a common method in speech synthesis evaluation. We used MOS (mean opinion score) and ranking to evaluate the performance. Initially, we select several audio samples related to the SVS task and categorize them into five groups based on audio quality, ranging from 1 to 5. These samples are not part of our training or test datasets and are used to familiarize participants with the scoring scale. We asked 23 participants to listen to synthesized audio samples along with ground truth, and rate them on a scale from 1 to 5 as part of the MOS evaluation. Additionally, we conduct a blinded ranking test. For each utterance ID, we present 8 audio samples, listed in \tabref{exp:main} in random order, and request participants to rank them based on audio quality. In this ranking, 1 indicates the highest quality, while 8 represents the lowest. A total of 126 test sets were presented to participants, with each participant randomly assigned 10 to 30 test sets for scoring and ranking, resulting in 464 evaluations. Each test set includes 8 audio samples requiring ranking and scoring (as shown in \tabref{exp:main}), along with 3 audio samples used for ablation studies (as shown in \tabref{exp:ablation}). We ensure that each test set is evaluated by at least 2 different participants.

Alongside subjective evaluations, we also utilize objective metrics to assess our model performance. The DNSMOS metric, proposed by \cite{reddy2022dnsmos}, is an objective, non-intrusive method for evaluating speech quality. DNSMOS employs deep learning models trained on extensive datasets to predict the subjective scores that human listeners would provide. We used two metrics from DNSMOS to evaluate our synthesized results: 1) SIG MOS, which measures the quality of the speech itself, independent of background noise, 2) BAK MOS, which assesses the intrusiveness and perceptibility of any residual background noise after noise suppression. Scores for both metrics range from 1 to 5, where 1 represents the lowest quality or most intrusive background noise, and 5 represents the highest quality or very suppressed background noise.

\subsubsection{Experimental Analysis}
\figref{4mel} provides a comparative analysis of the ground-truth mel-spectrogram and those generated by VISinger, DiffSinger, and RDSinger, all conditioned on the same music score. The mel-spectrogram generated by RDSinger (\figref{4mel}d) exhibits a higher level of detail compared to those produced by VISinger (\figref{4mel}b) and DiffSinger (\figref{4mel}c). The spectrogram from RDSinger closely resembles the ground-truth spectrogram (\figref{4mel}a), reflecting high fidelity in reproducing the original audio characteristics. Additionally, upon closer examination of the transition regions within the spectrograms, RDSinger demonstrates superior performance in preserving detailed acoustic features. Specifically, during pitch transitions and frequency transitions, RDSinger exhibits a more accurate and detailed representation compared to the other models.

The quantitative results of our experiment are detailed in \tabref{exp:main}. The reference quality, represented by GT-Mel-HiFiGAN, establishes an upper bound for the MOS with a score of 4.43. Our proposed RDSinger model achieves a MOS of 3.46, outperforming all other models in this study. Despite building our model upon the DiffSinger architecture, the introduction of the blur reference mechanism yields an improvement of 0.16 in MOS. Additionally, RDSinger achieves SIG MOS scores comparable to state-of-the-art models and shows better performance in background noise suppression, as evidenced by the significant improvement in BAK MOS. Ranking analysis shows that RDSinger achieved the highest ranking score among all evaluated models. \figref{people} illustrates the ranking distribution from all participants' evaluations, showing that most participants placed RDSinger in the 3rd or 4th position, indicating it was often considered among the top two models in comparisons. The substantial improvement in background noise quality and high rankings can be attributed to the reduction of artifacts in the synthesized audio, resulting in enhanced clarity and a better listening experience. These results affirm the efficacy of the enhancements introduced in RDSinger, particularly in maintaining high-quality voice synthesis.

\subsubsection{Ablation Studies}
We conducted several ablation studies to verify the effectiveness of each module in the proposed system. \tabref{exp:ablation} makes it evident that both subjective and objective metrics decline as components are removed from the model. Notably, the removal of the reference blur leads to a significant drop in model quality, with more unnatural artifacts occurring in pitch transition regions. This is because the reference mel-spectrogram, derived from FastSpeech2, has inherent difficulties with natural transitions, and using it directly as a reference further impairs the model's performance.
\begin{table}[h]
\centering
\begin{tabular}{l|ccc}
\toprule[2pt]
Method & MOS (↑) & BAK MOS (↑) \\
\hline
Full Model & \textbf{3.46} ± 0.07 & \textbf{4.07} ± 0.06 \\
W/o Reference Blur & 3.21 ± 0.09 & 3.96 ± 0.07 \\
W/o Weighted Loss & 3.40 ± 0.07 & 4.06 ± 0.06 \\
Only Reference Network & 3.29 ± 0.08 & 3.98 ± 0.06 \\
\bottomrule[2pt]
\end{tabular}
\caption{Ablation study on the different components of RDSinger}
\label{tab:exp:ablation}
\end{table}

Both RDSinger and DiffSinger employ diffusion denoising to generate audio. The number of denoising steps has a significant impact on model quality. As shown in \tabref{exp:steps}, our proposed model achieves optimal results with 100 steps. Remarkably, by directly incorporating the reference mel-spectrogram, RDSinger produces results comparable to DiffSinger’s 54-step model with just 24 steps.
\begin{table}[h]
\centering
\begin{tabular}{l|c|cc}
\toprule[2pt]
Method &  Steps & SIG MOS (↑) & BAK MOS (↑) \\
\hline
\multirow{2}{*}{DiffSinger} & 54 & 3.32 ± 0.06 & 3.85 ± 0.06 \\
                            & 100 & 3.39 ± 0.05 & 3.98 ± 0.06 \\
\hline
\multirow{3}{*}{RDSinger} & 24 & 3.27 ± 0.06 & 3.86 ± 0.06 \\
                          & 54 & 3.42 ± 0.06 & 4.04 ± 0.07 \\
                          & 100 & 3.47 ± 0.06 & 4.07 ± 0.06 \\
\bottomrule[2pt]
\end{tabular}
\caption{Comparison of Denoising Steps between DiffSinger and RDSinger}
\label{tab:exp:steps}
\end{table}

\section{Conclusions}
We introduce RDSinger, an SVS acoustic model based on a referenced diffusion probabilistic model. To improve conditioning accuracy, we developed a reference network to guide the denoising process. Noting misleading regions in the synthesized reference mel-spectrogram, we applied Gaussian blur to pitch transition areas and adjusted the loss weight. Experiments on the Opencpop dataset confirmed RDSinger’s effectiveness, showing improved synthesis quality and fewer artifacts compared to previous models.

\bibliography{aaai25}

\end{document}